# Raman spectra of $Ge_xAs_ySe_{100-x-y}$ glasses

S. Gapochenko[1], V. Belozertseva[1], H. Khlyap[2], A. Mamalui[1]

[1]Department of General and Experimental Physics, National Technical University «Kharkiv Polytechnic Institute», 21 Frunze St., Kharkiv 61002, Ukraine

[2]Distelstr.11, D-67657 Kaiserslautern, Germany

sdgapochenko@gmail.com

Raman spectra of $Ge_xAs_ySe_{100-x-y}$ ($0 \leq x \leq 30$; $10 \leq y \leq 40$) glasses have been studied at room temperature. Three sets of samples were investigated; it was revealed that they are qualitatively differing by shape of their spectra. The structure model was proposed for glasses of each set. The selection of the glasses on the sets with the same structure type can be made using the formal parameter ─ the mean coordination number $<m>$ (the mean number of covalent bonds per atom): i) $2.1 \leq <m> \leq 2.51$ ( polymeric structure); ii) $2.51 < <m> \leq 2.78$ (molecular-cluster structure); iii) $2.78 < <m> \leq 3$ (network structure).

## 1. Introduction

Chalcogenide glasses (CG) are widely used in technological applications such as infrared optical elements, acousto-optic and all-optical switching devices, holography recording media *etc* [1 ─ 6]. Information on the short-range order structure of chalcogenide glasses is particularly valuable in order to establish useful correlations between their structural and macroscopic properties [7, 8]. Moreover, the unique nature of glassy state is one of the most interesting and contradictory problems in solid state physics until recently. When studying glassy materials one faces with two principal difficulties. One is the absence of the general theory of a non-crystalline state, and the other is that the every traditional experimental probe of structural study "by itself does not yield a unique interpretation" [7]. As a consequence, the variety of structure models [8-17] were proposed for glassy materials. Sometimes they are in contradictory with each other (for example, Ref. [8] and [11]).

Chalcogenide glassy alloys are attractive for theoretical and experimental study. Being essentially covalent materials [11, 18], they may be served as the model objects. Just this feature of chalcogenide glasses made possible to develop the topological concepts of their structure [11, 12, 14, 19]. For example, the theoretical approach based on valence-force field approximation [19] had been fruitfully applied for description of the elastic properties of CG. The topological theories use the general parameter called the theoretical mean coordination number $<m>$ ($<m> = 2 + 0.02 \cdot x + 0.01 \cdot y$ for $Ge_xAs_ySe_{100-x-y}$ - like glasses), which governs the glass structure independent on atomic constituents. Including this parameter in the theoretical description of glasses is undoubtedly very useful, for the first time, because of appearing possibility of classifying the physical properties of glassy alloys differing in composition in the framework of one theoretical approach. Two topological phase transitions: from one-dimensional (1D) to two-dimensional (2D) structure at $<m>_{1c} \approx 2.4$, and further to three-dimensional (3D) one at $<m>_{2c} \approx 2.67$ were reported [14, 20, 21] for chalcogenide multicomponent glasses as a result of their experimental studies. At the same time, the authors [22, 23, 24, 25] have obtained another values for $<m>_{1c}$ and $<m>_{2c}$. The most interesting range of $<m>$ is $<m> \approx 2.4 \div 2.67$, where the very different dependencies of glasses physical parameters on <*m*> were observed even for different cross-sections of the same glass-forming system [20]. Earlier [26] we have studied the elastic properties of $Ge_xAs_ySe_{100-x-y}$ ($0 \leq x \leq 30$, $10 \leq y \leq 40$, $2.1 \leq$ <*m*> $\leq 3.0$) glassy alloys and revealed that there were three sets of glasses with different behavior of elastic parameters versus <*m*>. They were: i) <*m*> ≈ 2.1 ÷ 2.51; ii) <*m*> ≈ 2.51 ÷ 2.78; iii) <*m*> ≈ 2.78 ÷ 3.0. It was supposed that the structure of glasses of the first and the third sets is to be of 1D and 3D - type, correspondingly, and it did not essentially depend on the chemical composition. The situation for the glasses of the second set is more complicated. The analysis of our results and published works [27, 28] showed that the observed anomalies of elastic moduli dependencies on <*m*> could not be understand in the framework of insights about exclusively two-dimensional structure of glasses.

It is well-known that the infrared and Raman spectroscopic techniques are powerful tools for probing the structural properties of chalcogenide materials [29, 30]. For example, the analysis of Raman spectra of binary arsenic sulfide chalcogenide glasses, $As_xS_{100-x}$, evidenced the presence of phase separation effects for *x*<25 [31] and the occurrence of intrinsic nanoscale phase separation for *x* = 40 [32, 33].

## 2. Experimental

Chalcogenide glasses $Ge_xAs_ySe_{100-x-y}$ ($0 \leq x \leq 30$, $10 \leq y \leq 40$) (the compositions of alloys and their <*m*> are listed in Table 1 were synthesized by using elements (Ge, As, Se) of 6*N* purity, which were melted in evacuated (P~$10^{-5}$ Torr) and sealed silica ampoules at 950 °*C* for 24 hours and subsequently quenched in air. Raman spectra were measured for reflected light beam at room temperature. The samples were prepared as plane-parallel plates with optical quality surface. The standard back-scattering geometry of measurement was used. The laser line 630 nm of 25 mW *He-Ne* laser was used as a light source. Scattered light was analyzed using DFS-24 spectrophotometer. The measurements had been performed at split width 0.1 nm corresponding to the selected energy interval $\delta\varepsilon = 6.5 \times 10^{-4}$ eV. The velocity *V* of Raman spectra scanning was $V = 1.5 \times 10^{-4}$ eV/s, provided the exclusion of the time-dependent signal distortion because of $\delta\varepsilon/V\tau > 2$ ($\tau = 1$ s is a time-constant of recording system).

## 3. Results

Raman spectra of $Ge_xAs_ySe_{100-x-y}$ glasses under study are shown in Fig.1. One can see that the alloys can be arranged into three sets with similar Raman spectra shape. Their mean coordination numbers are: i) <*m*> ≈ 2.10 ÷ 2.40; ii) <*m*> ≈ 2.51 ÷ 2.78; iii) <*m*> ≈ 2.91 ÷ 3.00, respectively.

Raman spectra of first set of glasses (curves 4, 5 and 6) include a broad main band with maximum at $\nu \approx 250 cm^{-1}$ and a shoulder near $\nu \approx 244 cm^{-1}$. The origin of these spectral features for glasses studied will be discussed. There is also the less intensive peak at $\nu \approx 194 cm^{-1}$ in Raman spectra of Ge-containing glasses. Its intensity follows Ge content increasing (curves 5 - 8). It possibly arises from $A_1$ bond-stretching mode of $GeSe_4$ tetrahedron [35] (curve 3).

The most complex Raman spectra are observed for glasses of second set (curves 7, 8 and 9). For all the glasses the $194 cm^{-1}$ peak is the dominant spectral feature. The spectra peculiarities over the range of 280 ÷ 300 $cm^{-1}$ may be attributed to $\nu_3(F_2)$ stretching mode of $GeSe_4$ tetrahedron [35]. Another peak with intensity rising from $Ge_{13.5}As_{23.5}Se_{63}$ to $Ge_{22.5}As_{32.5}Se_{45}$ is observed at $\nu \approx 236 cm^{-1}$. Its origin is not clear in our case but some suppositions will be made. The vibration band at $\nu \approx 212$ cm$^{-1}$, appearing only for glasses of this set is of special interest. Its intensity corresponds to increasing Ge and As content. The so-called $A_1^c$ companion peak is observed in Raman spectra of glassy GeSe$_2$ [34] (curve 3), its origin appears as ambiguous [10, 11, 35]. The next vibration mode at $\nu \approx 130$ cm$^{-1}$ is not revealed in Raman spectra of glassy As$_x$Se$_{100-x-y}$ [10, 29], Ge$_y$Se$_{100-y}$ [10, 36] and (GeSe$_2$)$_x$(As$_2$Se$_3$)$_{100-x}$ [36]. However, it is inherent to the spectra of the third set alloys with essential content of Ge and As atoms. Raman spectra of Ge$_{27}$As$_{37}$Se$_{36}$ and Ge$_{30}$As$_{40}$Se$_{30}$ alloys (curves 10 and 11) demonstrate new spectral features in comparison with those of preceding sets. The peak at 182 cm$^{-1}$ (Ge$_{27}$As$_{37}$Se$_{36}$) which shifts to the long-wave region ($\nu \approx 187$ cm$^{-1}$) for Ge$_{30}$As$_{40}$Se$_{30}$ originates from the vibration of Ge-Ge bond in Ge$_2$Se$_{6/2}$ [37] (curve 3). One can see that peak width decreases for Ge$_{30}$As$_{40}$Se$_{30}$. The vibration band at $\nu \approx 116$ cm$^{-1}$ is also observed for crystalline GeSe$_2$ [37]. The bands over the region of $280 \div 300$ cm$^{-1}$ with weak-developed contour may be assigned to $\nu_3(F_2)$ vibration mode of GeSe$_4$ tetrahedron (curve 3). Our results principally correlate with data of [28] for As$_{50}$Se$_{50}$ doped with 1.0 at % Ge.

### 4. Discussion

The main band of glasses of the first set is similar to that of glassy Se (curve 1) [35]. In pure Se the 250 cm$^{-1}$ Raman band is assigned to $A_1$ and $E_2$ modes of the Se$_8$ ring molecule and a shoulder at 235 cm$^{-1}$ to the vibration modes of Se$_n$ polymeric chain [36]. The ratio of Se$_n$ chains to Se$_8$ rings was estimated [35] to be 0.23. Schöttmiller et al [36] have also reported that the 250 cm$^{-1}$ peak due to Se$_8$ molecules was still the dominant spectral feature for As$_x$Se$_{100-x}$ ($x \leq 16$) glassy alloys. In the framework of this model the observed peak-shoulder intensities ratio may be considered as an evidence of preferable Se$_8$ rings producing in examined glasses. At the same time, the peak at $\nu \approx 227$ cm$^{-1}$ was observed for Se-rich As$_x$Se$_{100-x}$ glassy materials [36]. This band was reported [36, 37] for glassy As$_2$Se$_3$ (curve 2) and is considered as assigned to $A_1$ stretching mode of AsSe$_3$ pyramidal unit. The band near $\nu \approx 227$ cm$^{-1}$ was attributed [36] to the growth of a polymeric network (AsSe$_{3/2}$)$_n$ at the expense of Se$_8$ molecules. As it follows from Fig.1 (curves 4, 5 and 6) the shoulder at 244 cm$^{-1}$ broadens with increasing As concentration. It is in agreement with the assumed growth of the number of AsSe$_{3/2}$ structural units.

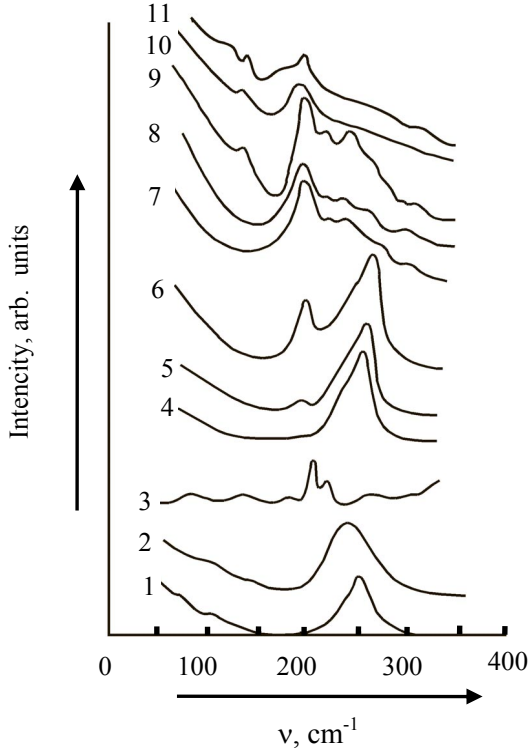

FIG.1: Raman spectra of $Ge_xAs_ySe_{100-x-y}$ glasses: 1 – $Se$; 2 – $As_2Se_3$; 3 – $GeSe_2$; 4 – $As_{10}Se_{90}$; 5 – $Ge_{4.5}As_{14.5}Se_{81}$; 6 – $Ge_{10}As_{20}Se_{70}$; 7 – $Ge_{13.5}As_{23.5}Se_{63}$; 8 – $Ge_{18}As_{28}Se_{54}$; 9 – $Ge_{22.5}As_{32.5}Se_{45}$; 10 – $Ge_{27}As_{37}Se_{36}$; 11 – $Ge_{30}As_{40}Se_{30}$

Thus, till now our results do not contradict to the above-mentioned insights [36-38] about structure of Se-rich $As_xSe_{100-x}$ glasses. However, for the studied $Ge_xAs_ySe_{100-x-y}$ samples Ge and As contents are monotonously increased (see Table 1). The 194 cm$^{-1}$ peak intensity follows Ge content increasing (curves 5 - 8). The Raman band at 201 cm$^{-1}$ was reported [39] for $GeSe_2$ and was assigned to the $A_1$ vibration mode of $GeSe_4$ tetrahedral unit. Therefore, we can suppose the formation of $GeSe_{4/2}$ structural units (s.u.) in investigated glasses. The simultaneous formation of $AsSe_{3/2}$ and $GeSe_{4/2}$ units must cause essential decreasing the number of $Se_8$ rings while increasing Ge and As content. This conclusion is not conformed to our experimental results (the position and the shape of the main band are almost not changed except its broadening with the change of glass compositions). Therefore, we can suppose that the glassy alloys of the first set have essentially polymeric structure. Polymeric chain $Se_n$ is their basic structural element. Branching $Se_n$ chains is promoted by Ge and As content increase. This conclusion is in agreement with our earlier experimental results [26] on elastic properties of studied glasses.

Raman spectra of glasses of the next set (curves 7, 8 and 9) show enough intricate shape. The spectral features at $\nu \approx$ 194 cm$^{-1}$ and ones over the range 280 ÷ 300 cm$^{-1}$ give an evidence for substantial density of $GeSe_{4/2}$ units. The peak at 236 cm$^{-1}$ may be attributed to $AsSe_3$ pyramidal unit (see curve 2). However, very close to its position, 230 cm$^{-1}$ band was observed [40] in Raman spectra of $Ge_xAs_2Se_3$ glasses and interpreted as corresponding to As-rich s.u. All studied glasses have an excess of As to Ge atoms. Besides this, there are not enough Se atoms for complete bonding of Ge and As atoms (Ge-Ge, Ge-As and As-As bonds must be produced) in the alloys of that set. Hence, in our case, 236 cm$^{-1}$ band may be also attributed to As-rich structural elements. Now we can not suggest more rigorous and detailed structure model of these As-rich elements because of the absence of more detailed experimental data. Our supposition about As-rich fragments being in glasses under investigation is confirmed by studying the redox chemical reactions in Ge-As-Se glasses [40]. Ge-compositions with low oxidation degree (GeSe) can reduce the As-based compositions. As a consequence of these processes, $GeSe_{4/2}$ s.u. must to be preferably formed. In Se-rich glasses $GeSe_{4/2}$ units are produced not at the expense $AsSe_{3/2}$ units. There are $GeSe_{4/2}$, $AsSe_{3/2}$ and $SeSe_{2/2}$ s.u. in alloys. The structure of glasses is essentially polymeric independently on the ratio of Ge and As concentration. As for materials with small deficit of Se these redox processes are preferable in producing the variety of different structures in dependence on components ratio.

The vibration band at 212 cm$^{-1}$ is of special interest. Its intensity increases proportionally to Ge and As concentrations. The same band is observed in Raman spectra of glassy $GeSe_2$ (see curve 3). There are some contradictory interpretations [10, 41, 11] of its origin.

Basing on the chemically ordered continuos random network (COCRN) model Lucovsky et al [10, 41] assumed that this band was due to splitting $\nu_1(A_1)$ mode of $GeSe_4$ as the result of bonding between $GeSe_4$ tetrahedrons. However, as was showed in [42] for crystalline $GeSe_2$ the angle between bonds at Se-atoms was equal to 100$^0$, so tetrahedral normal vibrations were weakly-bonded. Hence, the assumption about $\nu_1(A_1)$ mode splitting was not quite reasonable. The narrowness of that band ($\delta\nu_{1/2}/\nu$ < 0.015) is an evidence of enough high symmetry of corresponding structural unit. Therefore, it can not be part of "random" network. At the same time,

Phillips et al [32] analyzing the isotopic shift from $GeSe_2$ to $GeS_2$ supposed that the 212 cm$^{-1}$ band ($A_1^c$ is a companion line) was caused by chalcogen motion. Taking also in account the regularities in Raman spectra of $Ge_xSe_{100-x}$ glassy alloys in dependence on composition Phillips [11, 31, 42] suggested the structure model of glassy $GeSe_2$. The basic structural element is sufficiently partially polymerized cluster, which is a fragment of the high-temperature layer structure of crystalline $GeSe_2$. The characteristic feature of the cluster is the chalcogen dimers on its lateral edges. Their motion causes the 212 cm$^{-1}$ peak. This supposition is corroborated by theoretical stimulation of glassy $GeSe_2$ Raman spectra [42]. The experimental results obtained and above-mentioned arguments make real existence of the Ge-based clusters in the structure of studied glasses. Thus, the structure of glasses ($<m> \approx 2.51 \div 2.78$) includes Ge-based molecular clusters and As - rich structural groups. Their density is proportional to Ge and As content.

Spectra of $Ge_{27}As_{37}Se_{36}$ and $Ge_{30}As_{40}Se_{30}$ alloys (curves 10 and 11) have not any peculiarities which may be attributed to polymeric or molecular fragments of glasses of preceding sets. There are some spectral signs of Ge-Ge bond (the peak at $\nu \approx 187$ cm$^{-1}$), Ge-As bond (the peak at $\nu \approx 130$ cm$^{-1}$) and Ge-Se bond (the peculiarities over the region of $280 \div 300$ cm$^{-1}$). The proximity of s.u. vibration modes position and experimental difficulties in Raman spectra detection (the laser line is in the absorption region of glasses studied) may be a cause of the others bonds disappearance in spectra measured. These alloys are in the region in the phase diagram, where according to Feltz [40] the glass structure is essentially 3D. Basing on these arguments one can suppose 3D network structure for glasses of this set.

**5. Conclusion**

Basing on Raman spectra of $Ge_xAs_ySe_{100-x-y}$ ($0 \leq x \leq 30$; $10 \leq y \leq 40$) glasses we have determined that their structure type is dependent partially on mean coordination number. The alloys with $<m> \approx 2.1 \div 2.4$ exhibit the polymeric structure based on $Se_n$ chains which are cross-linked by Ge and As atoms. The basic structural units of this group of glasses are $SeSe_{2/2}$, $AsSe_{3/2}$ and $GeSe_{4/2}$. The structure of second set glasses ($<m> \approx 2.51 \div 2.78$) is derived from 2D Ge-based molecular clusters and As-rich structural fragments. Glassy $Ge_{27}As_{37}Se_{36}$ and $Ge_{30}As_{40}Se_{30}$ ($<m> \approx 2.91 \div 3.00$) compose the third set of alloys with 3D network.

**REFERENCES**

[1] E. Lepine, Z. Yang, Y. Gueguen, J. Troles, X.-H. Zhang, B. Bureau, C. Boussard-Pledel, J.-Ch. Sangleboeuf, and P. Lucas, J. Opt. Soc. Am. B **27**, (2010) 966.

[2] M.-L. Anne, J. Keirsse, V. Nazabal, K. Hyodo, S. Inoue, C. Boussard-Pledel, H. Lhermite, J. Charrier, K. Yanakata, O. Loreal, J. Le Person, F. Colas, C. Compère and B. Bureau, Sensors **9 (**2009) 7398.

[3] A. Zakery, S. R. Elliott, J. Non-Cryst. Solids **330** (2003) 1-12.

[4] A. R. Hilton, Chalcogenide Glasses for Infrared Optics, McGraw-Hill Professional, 2009.

[5] D. Lezal, J. Pedlikova, J. Zavadila, Chalcogenide Letters **1**(2004)1.

[6] M. Rozé, L. Calvez, J. Rollin, P. Gallais, J. Lonnoy, S. Ollivier, M. Guilloux-Viry, X. Zhang , Applied Physics A-Materials Science & Processing **98**, 1 (2010)97.

[7] L.Červinka, J. Non-Cryst. Solids **98**(1987)207.

[8] Ye. Porai-Koshitz, in: Fizika i Khimiya Stekla, (in Russian).Nauka, Moskva, 1987.

[9] M. R. Hoare, J.Baker, in: Structure of Non-Crystalline Materials,

P.H.Gaskell (Ed.),Taylor & Fransis, London, 1976.

[10] G.Lucovsky and R.M.Martin, J.Non-Cryst.Solids **8-10**(1972)185.

[11] J.C. Phillips, J.Non-Cryst.Solids **34**(1979)153;
J.C. Phillips, J.Non-Cryst.Solids **43**(1981)37;
J.C. Phillips, Phys.State Solids(b) **101**(1980)473.

[12] R.Zallen, The Physics of Amorphous Solids, Jhon Willey & Sons, N-Y, 1983.

[13] V.V. Tarasov, Problemy Fiziki Stekla (in Russian), Stroiizdat, Moskva, 1979.

[14] K. Tanaka, Phys. Rev. **B39**(1989)1270.

[15] A. Bakai, Poliklasternie amorfnie tela (in Russian), Energoatomizdat, Moskva, 1987.

[16] M. Popesku, A. Andriesh, V.Ciumash, M.Iovu, S.Shutov and D.Tsiulianu, Physics of Chalcogenide Glasses, Stiintza and Stiinfica Publ. Houses, Chishinau and Bucharest, 1996.

[17] O.Shpotyuk, J. Filipecki, Physics and Chemistry of Glasses - European Journal of Glass Science and Technology Part B. **46** 2 (2005)170.

[18] Z.U. Borisova, Khalcogenidnie Stecloobraznie Poluprovodniki (in Russian), Nauka, Leningrad, 1981.

[19] M.F. Thorpe, J.Non-Cryst.Solids **76**(1985)109.

[20] S. Mahidevan and A.Giridhar, J.Non-Cryst.Solids **110**(1989)118.
S. Mahidevan and A.Giridhar, J.Non-Cryst.Solids **143**(1992)52;

[21] J.Y. Duquesne and G.Bellessa, Europ.Lett. **9**(1989)453.

[22] R. P. Wang, D. Bulla, Anita Smith, T. Wang, and Barry Luther-Davies J. Appl. Phys. **109**, 023517 (2011).

[23] B.L. Halfpap and S.M. Lindsay, Phys.Rev.Lett. **57**(1986)847.

[24] Y. Ito and S. Kashida, Solid State Commun. **65**(1988)449.

[25] I. Fekeshgazi, K. May, Y. Mitsa and A.Vakareek, in Physics and Applications of Non-Crystalline Semiconductors in Optoelectronics, A.Andriesh and M.Bertolotti (Ed.), NATO ASI Series, 3.High Technologies – Vol.36, Kluwer Academic Publishers, Dordrecht/ Boston/ London, 1997.

[26] S. Gapochenko and V. Bazakutsa. J. Non-Crystalline Solids. **270** (2000)274.

[27] M. S. Iovu, S. D. Shutov, A. M. Andriesh, E. I. Kamitsos, C. P. E. Varsamis, D. Furniss, A. B. Seddon, M. Popescu. J. Optoelectron. Adv. Mater. 3(2)(2001) 443.

[28] M. S. Iovu*, E. I. Kamitsosa, C. P. E. Varsamisa, P. Boolchandb, M. Popescu. Chalcogenide Letters **2**(2005)21.

[29] E. I. Kamitsos, J. A. Kapoutsis, I. P. Culeac, M. S. Iovu. J. Phys. Chem. **B101** (1997) 11061.

[30] V. Kovanda, Mir. Vlček, H. Jain. J. of Non-Cryst. Solids **326&327**(2003)88.

[31] T. Wagner, S. O. Kasap, M. Vlček, A. Sklenař, A. Stronski. J. Non-Cryst. Solids **227&230**(1998) 752.

[32] D. G. Georgiev, P. Boolchand. Philosophical Magazine **83(25)**(2003) 2941.

[33] S. Mamedov, D. G. Georgiev, Tao Qu, P. Boolchand. J. Phys: Condens. Matter **15**(2003) 52397.

[34] G. Petrash, Optica i Spectroskopia (in Russian) **9**(1960)423.

[35] T. Mori, S. Onari and T. Arai, Jap.Appl.Phys. **19**(1980)1027.

[36] J. Schottmiller, M. Tabak, G. Lucovsky and A.Ward, J.Non-Cryst.Solids **4**(1970)80.

[37] G. Lucovsky, Phys.Rev.**B6** (1972) 1480.

[38] J. C. Griffits, G.P. Espinosa, J.C. Phillips and J.P. Remeika, Phys.Rev.**B28** (1983) 4444.
[39] C.H. Hurst and E.A.Davis, J.Non-Cryst.Solids **16**(1974)343.
[40] A. Feltz, Amorphous and Glass Inorganic Solid Materials (Russian traslation), Mir, Moskva, 1986.
[41] G. Lucovsky, R.J. Nemanich, S.A. Solin and R.C.Keezer, Solid State Commun. **17**(1975)1567.
[42] J.A. Aronovitz, J.R. Banar, M.A. Marcus and J.C. Phillips, Phys.Rev. **B28**(1983)4454.

Table 1. The compositions of the examined chalcogenide glassy $Ge_xAs_ySe_{100-x-y}$ $(0 \leq x \leq 30, 10 \leq y \leq 40)$ alloys and their mean coordination numbers $<m>$.

| N | Composition | $<m>$ | N | Composition | $<m>$ |
| --- | --- | --- | --- | --- | --- |
| 1 | $As_{10}Se_{90}$ | 2.10 | 5 | $Ge_{18}As_{28}Se_{54}$ | 2.64 |
| 2 | $Ge_{4.5}As_{14.5}Se_{81}$ | 2.24 | 6 | $Ge_{22.5}As_{32.5}Se_{45}$ | 2.78 |
| 3 | $Ge_{10}As_{20}Se_{70}$ | 2.40 | 7 | $Ge_{27}As_{37}Se_{36}$ | 2.91 |
| 4 | $Ge_{13.5}As_{23.5}Se_{63}$ | 2.51 | 8 | $Ge_{30}As_{40}Se_{30}$ | 3.00 |